\documentclass{article}

\usepackage{arxiv}

\usepackage[utf8]{inputenc} 
\usepackage[T1]{fontenc}    
\usepackage{hyperref}       
\usepackage{url}            
\usepackage{booktabs}       
\usepackage{amsfonts}       
\usepackage{nicefrac}       
\usepackage{microtype}      
\usepackage{lipsum}
\usepackage{graphicx}
\graphicspath{ {./images/} }
\usepackage[version=4]{mhchem}

\title{The Future of Computing for Materials Science Challenges}

\author{
  Phalgun Lolur\thanks{Corresponding author: \texttt{phalgun.lolur@capgemini.com}} \\
  Capgemini Quantum Lab \\
  147-151 Quai du Président Roosevelt \\
  92130 Issy-les-Moulineaux, France \\
  \And
  Richard P. Padbury \\
  Research Triangle Park \\
  Durham NC, USA \\
  \And
  George H. Booth \\
  Department of Physics, King's College London \\
  Strand, London WC2R 2LS, U.K. \\
  \And
  Katherine Inzani \\
  School of Chemistry, University of Nottingham \\
  Nottingham NG7 2RD, U.K. \\
  \And
  Nicole Holzmann \\
  PsiQuantum, 700 Hansen Way \\
  Palo Alto, CA 94304, USA \\
  \And
  Thomas W. Keal \\
  STFC Scientific Computing, Daresbury Laboratory \\
  Keckwick Lane, Daresbury, Warrington, WA4 4AD, UK \\
  \And
  Joseph Montoya \\
  Toyota Research Institute, 4440 El Camino Real \\
  Los Altos, CA 94022, USA \\
  \And
  Daniel F. Urban \\
  Fraunhofer Institute for Mechanics of Materials IWM \\
  Wöhlerstr. 11, 79108 Freiburg, Germany \\
  \And
  Thomas Eckl \\
  Robert Bosch GmbH, Robert-Bosch-Campus 1 \\
  71272 Renningen, Germany \\
  \And
  Emanuele Marsili \\
  Airbus Central Research and Technology \\
  Rond Point Maurice Bellonte, 1, 31707, Blagnac, France \\
  \And
  Wibe A. de Jong \\
  Lawrence Berkeley National Laboratory \\
  Berkeley, CA, USA \\
  \And
  Jonathan R. Owens \\
  GE Vernova Advanced Research, 1 Research Circle, Niskayuna, NY 12309 \\
  \& Ralph O'Connor Sustainable Energy Institute, Johns Hopkins University \\
  \And
  Julian van Velzen \\
  Capgemini Quantum Lab \\
  147-151 Quai du Président Roosevelt \\
  92130 Issy-les-Moulineaux, France \\
}

\begin{document}

\maketitle

\begin{abstract}
Materials discovery increasingly relies on the coordinated use of theory, computation, experiment, data-driven methods, and emerging quantum technologies, yet the full potential of these tools is realised only when they operate within workflows that reflect the complexity of real systems. This perspective summarises current capabilities, limitations, and opportunities across these domains, drawing on contributions from academia, industry, and national laboratories to identify the scientific and structural requirements for more reliable and efficient discovery. Classical simulations provide broad coverage across design spaces, while experimental measurements reveal degradation, heterogeneity, and kinetic processes that determine performance under realistic conditions. Machine learning accelerates exploration when supported by well-curated datasets with clear provenance and uncertainty quantification, and quantum computing offers promising routes into correlated electronic behaviour when aligned with properties that influence engineering decisions. Collectively, these insights highlight the need for reproducible workflows, shared data standards, realistic benchmarks, and a research culture that prepares scientists to work across paradigms. By integrating these methodological and organisational elements, the community can move toward discovery processes that deliver robust predictions, support confident decision making, and shorten the path from conceptual design to deployable materials.
\end{abstract}


\section{Navigating the Challenges of Materials Discovery}

The path from a promising idea to a material that can survive real industrial use has never been straightforward, and the distance between concept and deployment remains much larger than people outside the field tend to imagine. Discovery is not just a scientific exercise carried out on a clean theoretical canvas. It unfolds within the constraints of supply chains, processing routes, regulatory standards, environmental pressures, and the unforgiving realities of manufacturing lines that operate on an enormous scale. Those pressures combine to make the landscape of materials development one of the most demanding and least forgiving scientific arenas, particularly when a new material threatens to disrupt established infrastructure or requires changes that ripple throughout an entire industry.~\cite{Ref1_Olson2000, Ref2_Furrer2023}

\vspace{0.5em}

Computation has expanded the space of what can be imagined, yet the hopes generated by simulation often collide with the difficulties of industrial development. The challenge is not that the simulations are wrong, but that the systems we try to design have far more degrees of freedom than the abstract representations we build. The gap between theoretical performance and manufacturable reality therefore becomes the central obstacle, since industry cares about materials that work under real-world operating conditions, not materials that succeed only in the well-lit world of controlled models. Even where calculations are formally well defined, different approximations can disagree by tenths of an electron volt or more on the same system, which is often enough to reorder candidate materials in a screening pipeline.

\begin{figure}[t]
    \centering
    \includegraphics[width=\columnwidth]{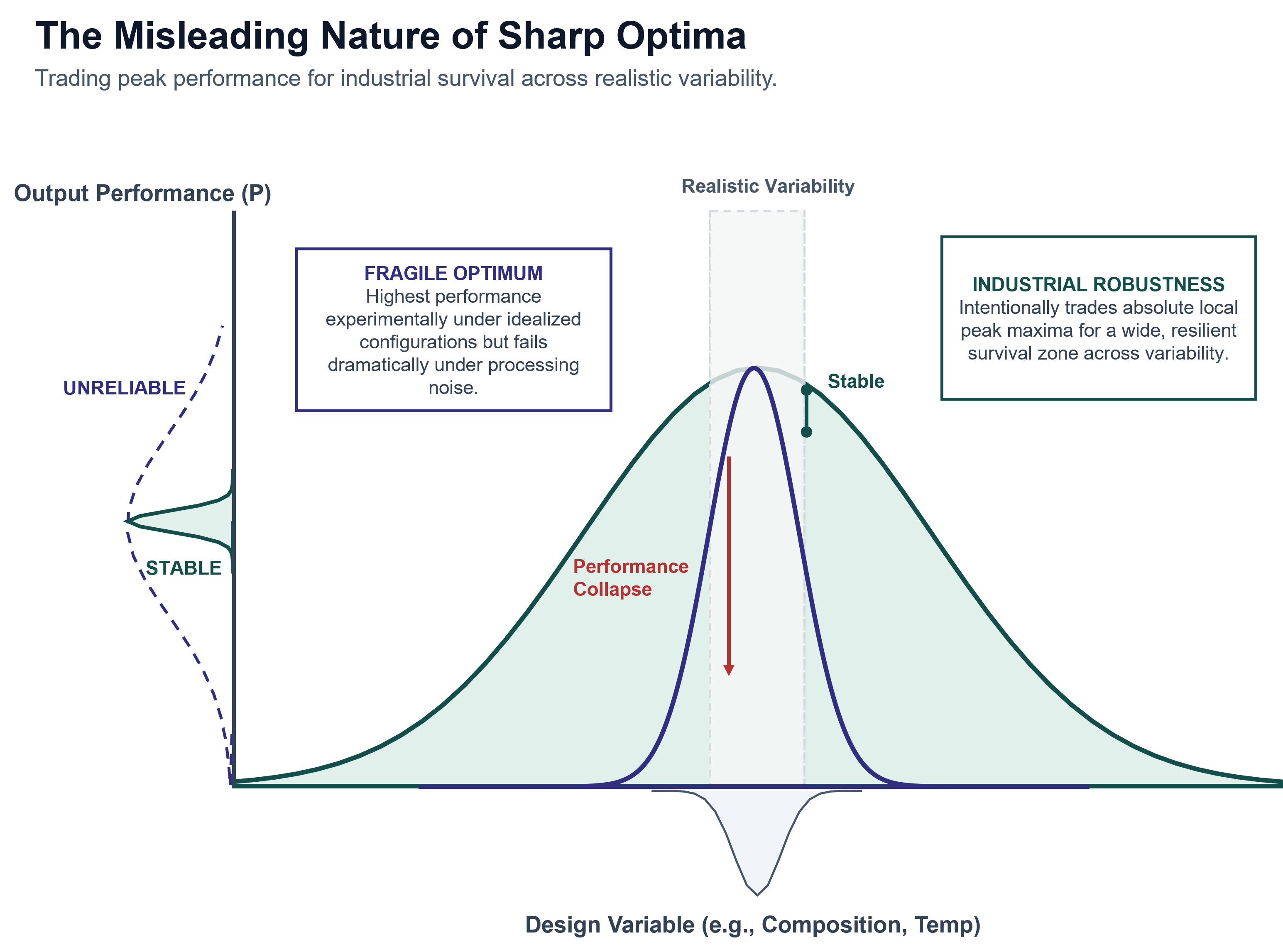}
    \caption{Trade-off between performance and robustness in materials design, where sharp optima are highly sensitive to variability, while flatter regions provide stable, reliable performance under realistic conditions.}
    \label{fig:uncertainty}
\end{figure}

\vspace{0.5em}

In sectors like aerospace, energy storage, automotive engineering, catalysis, and corrosion science, the same frustrating pattern regularly repeats. A new alloy survives laboratory testing, but accumulates microstructural defects during machining or heat treatment that were never anticipated during its design. A battery material offers exceptional gravimetric energy density yet fails long before a full cycle life is reached because the disorder present at manufacturing scale reshapes its degradation pathways. A catalyst that looks perfect from an electronic structure perspective becomes unusable because its active surfaces restructure under actual operating environments. The reality is that these situations do not sit at the edges of our field as rare curiosities but appear with such regularity that they define the practical boundaries of materials development whenever a candidate is pushed from controlled laboratory conditions into the complex and often unforgiving environments that accompany real industrial use.~\cite{Ref3_Wang2023, Ref4_Nian2023, Ref5_Ogley2025}

This is why many researchers have begun shifting emphasis away from chasing theoretical ``global optima'' and toward identifying ``robust optima''. A robust optimum is not the material with the highest predicted value on an idealised metric. It is the material that maintains its performance once the realities of processing variation, microstructural evolution, supply chain volatility, and operational uncertainty begin to push and pull its behaviour. That conceptual shift is subtle, yet it changes the workflow entirely. This shift is illustrated in Figure~\ref{fig:uncertainty}, where sharp optima give way to flatter more robust regions of the design space that better tolerate real-world variability. Robustness becomes an objective rather than an afterthought, and manufacturability becomes a design variable, not a late-stage constraint that forces unwelcome compromises long after enthusiasm has hardened around a particular candidate.

The uncomfortable truth is that many discovery pipelines still operate as if computation, synthesis, characterisation, processing, and scale-up exist in a sequence rather than a connected loop. In practice, every stage influences every other, and a workflow that isolates them will eventually fail to capture the interactions that determine real performance. This means the early phases of discovery must expand to include information that historically arrived only at the end. Processing windows, environmental limits, degradation mechanisms, lifetime constraints, and regulatory considerations all need to appear much earlier in the computational landscape. Without this, the models unintentionally search for materials that are attractive only in theory, while the materials that would succeed in practice remain hidden behind constraints the models never saw.

Industry, academia, and national laboratories approach these challenges with different instincts, shaped by their environments and responsibilities. Industrial researchers focus on manufacturability, safety, longevity, and stability across wildly varying operating conditions. Academic groups prioritise theoretical clarity, high-fidelity models, and fundamental insight. National laboratories contribute multiscale modelling, advanced characterisation tools, and the infrastructure required to test ideas under demanding conditions that mirror real operation. Although these three perspectives arise from different cultures, they converge on a shared recognition: materials discovery must integrate real-world constraints from the very beginning rather than treating them as filters applied after a computational search has produced a list of candidates.

By addressing these constraints early in the discovery loop, we view materials not as abstract chemical systems, but as engineered objects designed to perform reliably despite the inevitable fluctuations of large-scale manufacturing. This mindset does not diminish the value of theoretical elegance or computational sophistication. Instead, it gives researchers a more honest search space, one that reflects the true shape of the problem rather than an idealised projection of it. The goal is not simply to discover new materials but to discover materials that survive real-world environments. This remains the most demanding test any material will ever face.

\section{Exploring the Four Paradigms of Scientific Research.}

\begin{figure*}[t]
    \centering
    \includegraphics[width=0.95\textwidth]{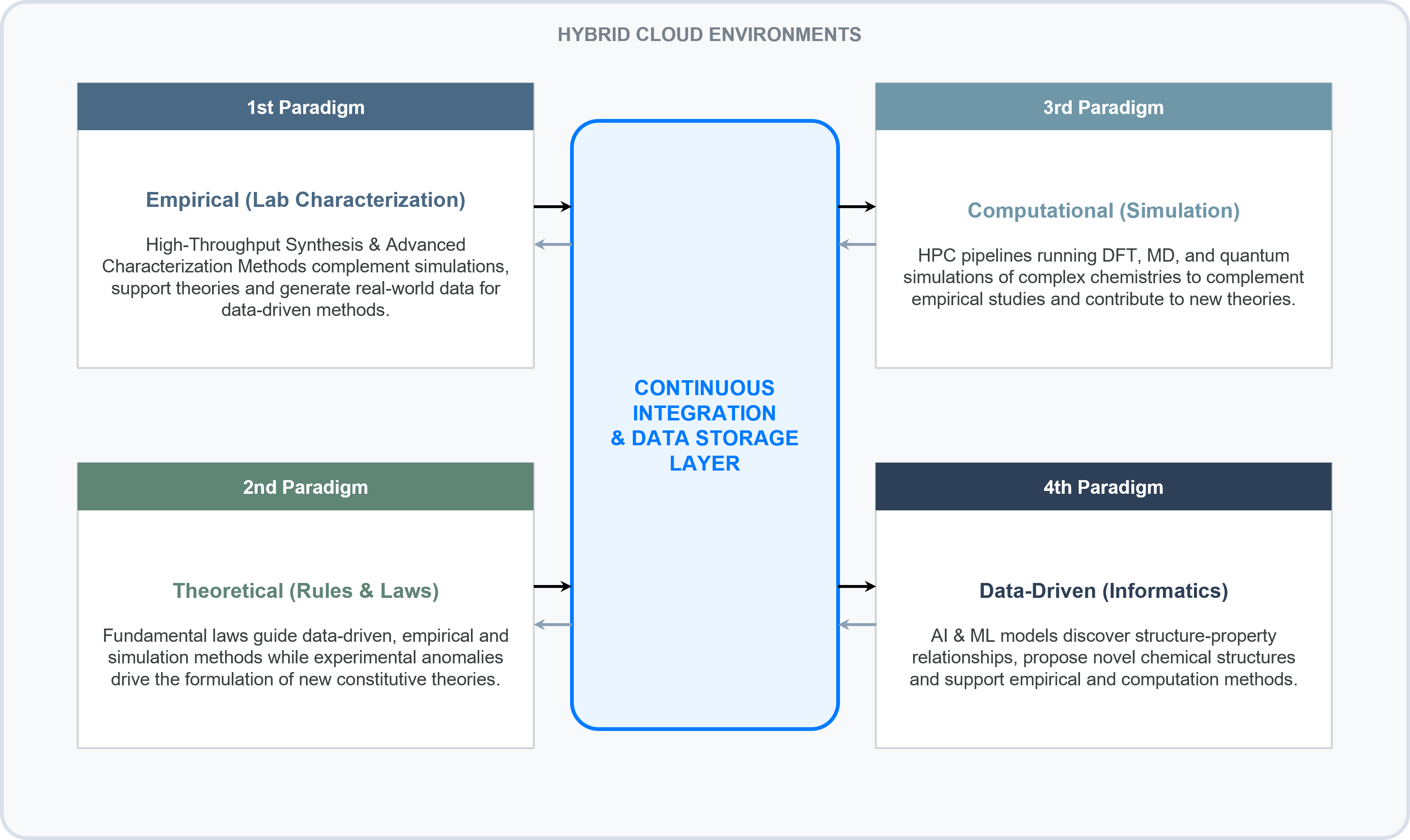}
    \caption{Evolution of scientific discovery from empirical and theoretical foundations to modern data-intensive paradigms enabled by advanced computing, hybrid cloud, and data technologies.}
    \label{fig:four_paradigms}
\end{figure*}

The evolution of scientific inquiry, as conceptualized by Jim Gray, has transitioned from traditional empirical and theoretical foundations into a sophisticated era of data-intensive discovery.~\cite{Ref6_Tolle2011, Ref7_Agrawal2016} This progression is also evident in materials discovery, which has evolved through successive paradigms, from direct observation to formal theory, then to computation, and now to data-driven inference---as shown in Figure~\ref{fig:four_paradigms}. These four paradigms do not follow a linear sequence, nor do they replace one another as the field evolves. Rather, they act as overlapping lenses that sharpen different facets of the same problem. The aim is not to emphasize the paradigms individually, but to establish workflows that enable the rapid exchange of evidence and interpretation between them. The most meaningful advances emerge when these approaches are permitted to interact without rigid disciplinary boundaries dictating which questions may be pursued.

The empirical tradition remains the foundation because materials ultimately reveal themselves through behaviour rather than through theory alone. Experimentalists uncover unexpected phases, hidden transitions, kinetic bottlenecks, metastable states, and degradation pathways that theory often fails to anticipate. These observations correct our assumptions about how matter behaves once it escapes the idealised world of controlled variables and exact boundary conditions. Empirical observation also brings an awareness of scale, heterogeneity, and imperfection, all of which matter tremendously in real manufacturing environments.

Theory enters not as decoration but as the structure that gives experiments meaning. Theoretical frameworks establish the rules that govern electronic structure, bonding, thermodynamics, reaction kinetics, microstructural evolution, and long-range interactions. When done well, theory offers a language that links phenomena across time and length scales. It allows researchers to recognise when two apparently unrelated behaviours share a common origin. However, theory also carries blind spots. It often relies on assumptions about locality, symmetry, or separability that break when confronted with messy systems containing defects, interfaces, disorder, or correlated electrons.

Computation transformed the landscape by giving researchers the ability to explore possibilities far beyond what could be synthesised or measured directly. Electronic structure methods, atomistic simulations, mesoscale models, and continuum descriptions each illuminate different parts of the materials space. Yet every computational method inherits biases from the physical assumptions on which it is built. Density functional theory performs well for many systems but struggles with strongly correlated materials, charge-transfer phenomena, rare-earth chemistry, and complex interfaces. Machine-learned potentials expand the accessible scale but inherit the limitations of the electronic structure data used in their training. A computational workflow gains power not by relying on a single method but by using a spectrum of approaches that help expose each other's weaknesses.

The rapid rise of data-driven methods added a new horizontal layer that links the vertical pillars of experiment, theory, and simulation. Machine learning can extract patterns from large and heterogeneous datasets that defy intuitive understanding. It reveals underlying structure in high-dimensional spaces, uncovers design principles that cross material classes, and identifies promising candidates in regions of the search space that human intuition tends to overlook. At the same time, data-driven methods are not magic. They require careful curation, physically meaningful descriptors, rigorous uncertainty quantification, and continuous integration with experiments and simulations so that the models remain grounded in genuine physical behaviour rather than statistical artefacts.

The most transformative developments occur when these paradigms cease functioning as isolated tools and instead operate in a linked cycle. An experiment uncovers an unexpected phase, prompting new theoretical interpretation. That interpretation suggests a computational approach capable of modelling the behaviour more accurately. The simulation generates structured data that feeds a machine-learning model, which then identifies trends that guide the next experiment. These feedback loops can collapse the time required to move from idea to insight because each paradigm enriches the others rather than waiting in line for its turn. A good example is one where these modes are not stacked in sequence, but are coupled tightly enough that each step changes what the next step does.

Battery research provides a clear illustration of the cycle.~\cite{Ref8_Kusne2020, Ref9_Lynch2009} Electronic structure calculations can rank candidate cathodes by redox chemistry, diffusion barriers, and interfacial energetics, but those predictions often break once particle cracking, surface reconstruction, and side reactions begin to dominate behaviour under cycling. High-throughput experiments expose those failure modes directly through capacity fade, impedance growth, and shifts in structural signatures over time. Machine-learning models then link early-cycle signals in voltage curves and diagnostic measurements to later degradation and lifetime variability. Theory and simulation return at that point to test competing mechanisms, not to decorate the story. The loop becomes useful because it forces disagreement into the open early, when it is still cheap to change direction.~\cite{Montoya2024}

The implication is organisational as much as technical, because these feedback loops only function when teams are built to operate across boundaries. Academic groups, industry researchers, and national laboratories each feel the implications of this shift in different ways. Universities are adapting graduate programmes to teach students how to collaborate across boundaries that used to be sharply defined. Industrial Research and Development teams are restructuring workflows so that theory, computation, and experiment operate in a continuous loop rather than a series of handoffs. National laboratories are developing shared computational platforms, high-throughput experimental capabilities, and data infrastructures that allow researchers from different sectors to engage with the four paradigms in unified workflows.~\cite{Abramov2025, Szymanski2023}

The field appears to be moving toward a model where the distinctions between empirical, theoretical, computational, and data-driven approaches matter less than the quality of the connections between them.~\cite{Ref10_dePablo2019} Materials discovery no longer progresses through a linear pipeline but through a network of interacting paradigms that reinforce one another. The task ahead is not to perfect any single approach but to make the movement between them as natural as possible, since this movement is where insight accumulates fastest and where the most resilient discoveries are made.

\section{Tackling the Early Stages of Discovery}

The earliest phase of discovery starts with a deceptively simple question that hides a very large problem, namely where to look first in a search space that grows faster than intuition can follow. Chemical composition, crystal structure, microstructure, processing history, and operating environment all conspire to create a landscape that cannot be sampled exhaustively, even with generous computing budgets and well organised experimental campaigns. Useful progress therefore depends on strategies that narrow attention without collapsing diversity too early, since premature focus routinely blinds projects to regions where the most interesting behaviour actually lives.

Inverse design has changed how many groups think about this first step because it reverses the traditional flow of reasoning in a way that suits decision making.~\cite{Ref11_Zunger2018} Rather than picking a material and asking what properties it might show, researchers specify the properties that matter for a particular application and then ask which material families could plausibly deliver them under realistic constraints. That reversal forces the pipeline to carry manufacturability, safety, regulatory limits, and supply chain stability alongside the usual performance metrics, which produces candidate sets that look less glamorous on paper but survive the realities of industrial development with far higher probability.

No single computational method provides a safe map across unfamiliar chemistry, which becomes painfully clear whenever correlation, disorder, strong coupling, or complex interfaces control behaviour of the materials to be designed. Density functional theory still anchors many workflows because it balances cost and accuracy, yet it carries biases from its functional choices and training habits that become visible whenever predictions leave well explored bonding regimes. Machine-learned interatomic potentials extend our reach into larger systems and longer timescales, although they inherit any weaknesses present in their training data and can struggle when confronted with conditions that differ from their formative experience.~\cite{Ref12_Behler2007} The practical lesson is simple enough to state and hard to execute, because reliable search requires methodological diversity, calibrated uncertainty, and honest escalation to higher fidelity when lower levels start to drift.

Exploration needs principled steering rather than heroic sampling, which is why active learning, evolutionary heuristics, and Bayesian optimisation have become so prominent in early discovery. These methods help decide where an additional calculation or experiment will actually reduce uncertainty in a way that matters for downstream choices, rather than simply inflating a dataset without improving judgment.~\cite{Ref13_Oganov2019, Ref14_Lookman2019, Ref15_Oganov2006, Ref16_Woodley2008} Dimensionality reduction and physically motivated descriptors reveal structure in high-dimensional spaces that otherwise look featureless, allowing search policies to move efficiently without losing sight of mechanistic insight. When these tools sit in a loop with targeted experiments, they transform exploration from a scattershot survey into an organised conversation with the system.

Early predictions rarely match full scale behaviour one to one, especially in materials where disorder and heterogeneous environments dominate the measured response. Computation still earns its place by acting as a compass rather than a street map, since consistent trends across approximate models can guide formulation, processing choices, and screening priorities even when point predictions remain imperfect. That guidance becomes particularly valuable in classes like sorbents, electrolytes, and corrosion inhibitors, where the relevant chemical space is enormous and the properties depend on mixtures, interfaces, and service environments that defy exhaustive enumeration.

Industrial programmes have more objectives than any isolated algorithm wants to handle, which means multi-objective thinking needs to appear at the start rather than the end. Toxicity, flammability, stability windows, supply chain security, recyclability, and cost impose boundaries that are as real as the electrochemical or mechanical targets that usually headline proposals. Treating those boundaries as design variables rather than after-the-fact filters produces candidate sets that look conservative yet move faster through development, because the trade-offs were acknowledged early and encoded directly into the search.

Data quality decides whether any of this machinery helps or hinders progress, since poor context quietly destroys comparability across experiments, simulations, and plants. Metadata about sample history, processing conditions, uncertainty estimates, and measurement protocols must travel with every datum if models are expected to generalise beyond the narrow conditions that produced them. Closed loops that join simulation, lab work, and analysis only function when the data backbone is clean and consistent, otherwise the loop chases ghosts and reinforces its own biases. In early discovery, fewer points with high information value beat large volumes of loosely annotated numbers every single time.

High performance computing remains essential because early exploration still demands sweeps across structures, compositions, and conditions that cannot be covered on modest resources. Shared facilities and coordinated programmes help teams modernise codes, adopt new hardware, and implement workflows that combine electronic structure, atomistic dynamics, mesoscale modelling, and surrogate models in a way that keeps throughput high without disconnecting from physics. As these infrastructures mature, they also enable reproducibility through shared environments and reference workflows, which matters greatly when many groups are exploring similar chemical territories with slightly different toolchains.

The role of experiment in this phase is changing from confirmation to interrogation, because the fastest progress comes when measurements are chosen to break degeneracies that models cannot resolve. Small, well targeted experiments that expose rate-limiting mechanisms, surface reconstruction, microstructural evolution, or stability envelopes can redirect entire campaigns before significant resources are committed.~\cite{Ref17_Rohr2020} When experimental design is informed by model uncertainty rather than habit or convenience, discovery accelerates not by doing more but by learning faster with less.

Putting these pieces together yields a first-stage workflow that behaves more like a learning system than a sequence of gates. Properties that matter to engineering are defined up front, constraints that shape industrial viability are placed alongside them, and search strategies are chosen to surface promising regions quickly while admitting when confidence is low. Computation proposes, experiments interrogate, data infrastructure remembers, and the cycle repeats until a small set of candidates survives both the physics and the pragmatics. Those candidates then earn the right to face the harder stages of development, which is where the next sections of the manuscript will focus.

\section{Harnessing Advancements in Computational Methods}

Modern computational methods have expanded far beyond their early role as tools for isolated property predictions. They now shape how researchers think about materials, how they organise workflows, and how they make decisions under uncertainty. What distinguishes recent advancements is not only the steady improvement in accuracy but the shift toward approaches that acknowledge the full complexity of real materials, including the effects of temperature, disorder, microstructure, and the operational environments that determine long-term performance. These developments are reshaping both the questions researchers ask and the kinds of answers they expect computation to deliver.

Density functional theory still anchors a large part of our computational infrastructure because it provides a workable compromise between physical fidelity and computational cost. Over many years researchers have extended its reach through new functionals, improved pseudopotentials, and more robust numerical strategies.~\cite{Ref18_Teale2022} Even so, its limitations remain clear whenever strongly correlated behaviour, charge-transfer phenomena, heavy elements, complex defects, or chemically diverse environments determine performance. These gaps become increasingly visible as materials challenges push into spaces where traditional approximations knowingly stretch beyond their comfort zones. The SCAN family and other modern functionals represent meaningful improvements, yet the field continues to encounter classes of systems where accuracy remains elusive even with significant computational effort.~\cite{Ref19_Sun2015}

One of the most important conceptual advances has been the acceptance that uncertainty matters as much as accuracy.~\cite{Ref20_Honarmandi2020} A single predicted value means very little without an understanding of the sensitivity, error propagation, and context that shape its reliability. Researchers now expect models to report confidence intervals, sensitivity to initial conditions, and the degree to which the underlying assumptions might bias outcomes. This shift has encouraged the development of internal validation strategies, cross-method comparisons, and practical definitions of adequacy tied directly to downstream decisions rather than to abstract benchmarks detached from use. When a simulated quantity drives an engineering choice, the difference between a tolerable approximation and a damaging misprediction depends on a clear sense of what the model actually knows and what it is merely guessing.

Multiscale modelling has become increasingly important because many critical properties emerge only when behaviour across several scales is connected in a physically meaningful way.~\cite{Ref21_Fonte2026} Electronic structure calculations reveal the energetic landscape in which atoms and electrons interact, but they are challenging to scale to complex systems and cannot easily describe longer timescale processes such as grain growth, defect mobility, crack initiation, or longer timescale diffusion on their own. Atomistic molecular dynamics captures some of these processes, although it struggles with the timescales and system sizes that define macroscopic environments. Mesoscale and continuum models describe microstructure evolution and macroscopic performance, yet require accurate inputs from lower levels to remain predictive. The challenge lies not in having these models available but in coupling them without losing mechanistic fidelity, numerical stability, or a clear interpretation of what information should flow from one scale to the next.~\cite{Ref22_Fish2021} Hybrid quantum/classical approaches have been successfully applied to materials challenges where an electronic structure description is crucial,~\cite{Ref23_Lipparini2021, Ref24_Lu2023} but further work remains to be done to connect all relevant scales.

Machine learning plays a growing role in both extending and connecting these scales. It can help create interatomic potentials with accuracy approaching quantum mechanical methods, track uncertainty through complex workflows, and bridge the hierarchy of models by distilling essential features from detailed simulations. These capabilities enable simulations that once required months of computation to run at speeds compatible with high-throughput searches. At the same time, they introduce new responsibilities because any machine-learned model inherits the assumptions, gaps, and biases of the data used for training. Physical validity can be compromised if models are not constrained by symmetries, conservation laws, or known limits, which is why physics-informed architectures and careful curation of training data have become essential ingredients in modern computational pipelines.

For materials where environmental conditions dictate performance, computation increasingly targets behaviour under finite temperature, variable pressure, chemical gradients, and evolving surface states. Static ground-state calculations, while useful, can miss the mechanisms that determine corrosion resistance, catalytic activity, mechanical reliability, or electrochemical stability. Researchers are now building workflows that combine electronic structure, reactive molecular dynamics, thermodynamic integration, and kinetic modelling into cohesive frameworks capable of predicting how a material evolves under load, cycling, humidity, or chemical exposure. This evolution from static snapshots to dynamic descriptions represents a critical shift because it aligns simulations with the conditions that engineers actually care about when qualifying materials for industrial deployment.

Facility-scale computing and coordinated software modernisation efforts are enabling these advances to reach broader communities. As codes adapt to heterogeneous hardware, GPU-accelerated infrastructures, and exascale architectures,~\cite{Ref25_Keal2022} simulations that once belonged to specialised groups now become accessible to a wider audience. This broadening of access matters because it allows more groups to participate in high-fidelity modelling, cross-validate results, and build shared benchmarks that reflect real materials challenges rather than idealised test sets. National laboratories, in particular, are creating environments where simulation, analysis, and data management operate as a single ecosystem rather than as separate activities.

The most valuable computational advances are those that improve decision making rather than those that merely sharpen numerical accuracy. A method that produces a more accurate prediction may offer little practical value if the improvement does not influence which candidate should advance or which experiment should be conducted next. Conversely, a method that provides transparent uncertainty quantification, improved interpretability, or better coupling across scales can transform an entire discovery pipeline even if its raw accuracy remains unchanged. Computational materials science is therefore moving toward an approach where algorithms are judged not simply by how closely they reproduce known numbers but by how effectively they guide exploration, reduce avoidable risk, and provide insight into the mechanisms that shape material behaviour.

These advancements signal a transition from computation as an isolated capability to computation as an organising principle for discovery. High-fidelity methods, uncertainty-aware frameworks, multiscale models, and machine-learned tools now operate alongside one another in workflows that mirror the complexity of the materials they aim to describe. The challenge ahead lies not only in continuing to improve accuracy but in building the integrated computational environments that allow researchers to navigate the expanding landscape of materials science with confidence, creativity, and a clear understanding of where the models remain reliable and where they begin to drift.

\section{Leveraging Artificial Intelligence and Machine Learning}

Artificial intelligence entered materials science promising speed, scale, and a fresh angle on structure--property relationships, yet the places where it truly changes outcomes are the places where it helps researchers decide what to do next rather than the places where it merely predicts one more number with slightly lower error.~\cite{Ref26_Butler2018} The value shows up when models reveal patterns that cut across composition, microstructure, and operating history, when those patterns can be explained in the language of mechanisms rather than black-box correlations, and when the same models drive an experiment or a simulation that would not otherwise have been run. That shift from passive prediction to active guidance is the thread that connects the most credible uses of machine learning today, whether the context involves screening catalysts, ranking corrosion inhibitors, triaging electrolyte formulations, or steering experimental campaigns in batteries and magnets.

The strongest projects begin with a simple acknowledgement that materials science rarely suffers from a lack of models and usually suffers from a lack of the right data, because the phenomena that matter most tend to depend on history, defects, interfaces, and environments that do not sit neatly in a single table of properties.~\cite{Ref27_Ramprasad2017, Ref28_Batra2020} Rather than treating data volume as a universal solvent, successful teams invest early in curation, standard operating procedures, robust metadata, and uncertainty annotations that capture how a datum was produced and how far it can be pushed before it breaks. Once this backbone exists, even modest datasets can support high-leverage learning with classical techniques that behave well in small-data regimes, while larger programmes can layer modern architectures that exploit structure in images, spectra, time series, and graphs without disconnecting from physics.

Physics-informed approaches deserve the attention they are receiving, not because they are fashionable but because they anchor learning to constraints that survive changes in composition, scale, and operating envelope.~\cite{Ref29_Karniadakis2021} Conservation laws, symmetry, invariances, and mechanistic priors reduce the space of plausible models to the subset that respects how matter behaves, which lowers data requirements and improves generalisation when the next material or device does not look exactly like the last. When these constraints are visible and testable rather than merely implied, they also improve trust, because engineers can interrogate sensitivities and trade-offs with the same vocabulary they use for process windows, safety margins, and lifetime predictions.

Interpretability is not a luxury that can be added at the end of a project through a post-hoc plot or a feature-importance bar chart, since it determines whether a model will be used in the parts of the organisation where decisions carry risk and accountability.~\cite{Ref30_Xie2018, Ref31_Reiser2022} Models that expose their uncertainty, document the limits of their training distribution, and map their inputs to intermediate physical quantities create a path from a recommendation to an action that engineers and scientists can defend. In discovery settings this means a ranking or a suggested next experiment arrives with a rationale tied to recognisable descriptors or mechanisms, while in development settings it means a forecast of performance includes confidence intervals and failure modes that drive targeted validation rather than vague reassurance.

In batteries and other time-dependent systems, the most effective pipelines do not attempt to learn directly from raw curves with minimal structure, because unstructured time series invite spurious patterns and brittle extrapolations. Teams that perform well compress signals into physically motivated summaries, such as features capturing early-life impedance growth, voltage hysteresis, plateau evolution, or relaxation signatures that reveal kinetic bottlenecks, then allow semi-supervised or supervised models to learn relationships to state of health, cycle life, or abuse tolerance. That combination of human-meaningful features and data-driven mapping travels better across chemistries and protocols, and it gives researchers levers they can actually pull when a model points to a weakness that would otherwise remain hidden behind a good overall fit.~\cite{Ref32_Oviedo2019}

Active learning closes the loop by turning models into instruments that choose what to learn next, which matters because the most expensive part of many programmes is not training a network but paying for the next calculation, synthesis, or characterisation. When the selection policy is tuned to reduce decision-critical uncertainty rather than to chase marginal accuracy on already familiar regions, campaigns converge faster on candidates that deserve scarce experimental time, and dead ends are pruned before they accumulate cost.~\cite{Ref33_Rohr2020Duplicate} The same principle extends to simulation, where uncertainty-aware surrogates can decide which expensive electronic-structure calculation will most improve a mesoscale model, or which microstructural scenario will most clarify a degradation pathway that affects lifetime.

Generative models have a place, although their usefulness depends less on raw creativity and more on whether the labels that drive selection correspond to realities that persist outside idealised conditions. In molecular design, where toxicity, volatility, stability, formulation compatibility, and cost compete with performance, a generative model trained on sparse or inconsistent annotations can fill a candidate list with elegant structures that fail immediately once process and safety constraints appear. The path forward involves coupling generative proposals with simulation and targeted testing that produce the dynamic labels which actually govern behaviour in mixtures, at interfaces, or under service environments, because static labels alone cannot represent the physics that determines whether a candidate advances beyond a controlled bench-top setup.

Industrial datasets present a different challenge, since they often contain decades of valuable history scattered across incompatible formats with missing context, partial measurements, and undocumented protocol drift. Progress in these settings begins with consolidation and standardisation that bring older records into a shared schema with common units, controlled vocabularies, provenance, and audit trails that record transformations, after which classical methods like kernel regression, support vector machines, and Gaussian processes often outperform heavyweight architectures by providing calibrated uncertainty and transparent behaviour in small to medium data regimes. As the backbone strengthens and targeted data generation fills the most damaging gaps, more expressive models can be introduced without sacrificing reliability, because they now stand on a foundation that encodes how and where their predictions can be trusted.

The role of national facilities and large research programmes is pivotal because they can create reference datasets and benchmark tasks that the community actually wants to use, not because they are large but because they are carefully designed to probe mechanisms that industry and academia both consider decision-relevant. When experimental ontologies, measurement protocols, and metadata standards are defined with input from practitioners who build products, those datasets become living resources that support reproducible comparisons across algorithms and toolchains, and they cut months from the time required to set up a credible pipeline in a new organisation or domain. Shared benchmarks also discourage the unhelpful habit of optimising models for narrow leaderboards detached from physics, since the metrics reward uncertainty handling, robustness under shift, and the ability to translate into experiments that change understanding rather than into fashionable curves that impress only at first glance.

Closing the loop between simulation, machine learning, and experiment is where momentum accumulates, because each step generates information that the others can use immediately rather than after long pauses for human arbitration. Electronic-structure calculations can produce targeted data to train surrogates that explore design spaces quickly, while those same surrogates can recommend the next high-value calculation that clarifies a mechanism in doubt.~\cite{Gharakhanyan2026} Laboratory platforms, increasingly automated and instrumented, can execute minimal experiments that remove the largest uncertainties in a model-guided plan, after which results flow back into training with full provenance and uncertainty estimates that reflect instrument limits and protocol variation. In this setting, a model is not a final answer but a participant in a conversation that moves continuously toward candidates that are not only novel and performant but also manufacturable and safe.

Safety, compliance, and sustainability enter earlier when machine learning participates in decision making rather than after the fact, because the same pipelines that predict performance can be trained to respect exclusion lists, regulatory thresholds, lifecycle constraints, and resource criticality that shape whether a material can be produced at scale. When these constraints live inside the optimisation rather than at the end of a funnel, the search space tilts toward candidates that survive scrutiny rather than toward candidates that will later require compromises that erase the apparent gains of early performance. This approach feels conservative on paper and proves faster in practice, since projects spend less time polishing materials that will never pass a gate that was visible from the start.

The surface area where AI meets people deserves deliberate design, because adoption depends on workflows that allow engineers and scientists to interrogate, adapt, and override model recommendations without friction. Dashboards that expose uncertainty, sensitivity to inputs, and predictions under counterfactual scenarios help teams test the consequences of design choices before they commit to expensive experiments or scale-up trials, while model registries and governance processes ensure that versions, training data, and validation results are traceable across projects and audits. These basic practices turn machine learning from an occasional curiosity into infrastructure that teams trust enough to use when it counts.

As AI takes on a larger role, the goal is not to replace simulation or experiment but to help them work together with a clarity and pace that would be impossible without a common memory and a set of principled shortcuts. The most durable wins arrive when learning reduces uncertainty in the places where uncertainty blocks progress, when it translates diverse signals into choices that researchers understand, and when it makes discovery feel less like wandering and more like purposeful movement through a complicated landscape that still rewards careful judgment.

\section{The Indispensable Role of Physical Experimentation}

However sophisticated our simulations or machine-learning pipelines become, materials ultimately prove themselves under measurement, and the experiments that matter most are rarely the ones that simply confirm what a model already expects. Experiments reveal misfit between tidy assumptions and unruly reality, exposing microstructure, interfaces, impurities, kinetic traps, and environmental sensitivities that remain invisible when we focus only on idealised structures and controlled boundary conditions.~\cite{Ref34_HattrickSimpers2016, Ref35_Gregoire2023} When researchers treat measurement as a genuine probe rather than a box-checking step, models learn faster, design spaces tighten intelligently, and programs spend less time polishing candidates that look impressive in theory yet crumble under realistic operating envelopes.

The right relationship between experiment and computation is not adversarial, and it is not hierarchical either, because neither side consistently outruns the other across problems that truly matter. Computation proposes mechanisms, suggests regimes to explore, and surfaces trends that steer attention toward fruitful regions of composition, structure, and process, while targeted experiments interrogate those proposals with conditions that reflect manufacturing practice and service life. That interplay turns data into understanding, since discrepancies become sources of insight rather than sources of embarrassment, revealing which approximations carry too much weight and which operating variables deserve more careful control.

Designing experiments to break degeneracies pays greater dividends than designing experiments to collect volume, particularly during the stages when uncertainty blocks decisions more effectively than missing accuracy. If two explanations fit the same property curve, the fastest progress comes from a measurement that separates those hypotheses with minimal cost, rather than from an extensive campaign that gathers redundant evidence around an already crowded point. Active learning can help here, because uncertainty-aware policies can propose the next experiment that is most likely to reduce decision-critical ignorance, whether the target involves surface reconstruction, metastable phase formation, defect mobility, electrolyte degradation, or stress-induced transformations that only appear after realistic cycling histories.

No realistic device contains the perfect material described in a ground-state calculation, and many of the properties that limit performance depend on microstructure, disorder, and rare events that unfold slowly under load. Lifetime emerges from crack initiation, creep, fatigue, corrosion, phase separation, and chemical drift, each influenced by histories that are difficult to encode a priori. Structured experimental programs that capture these slow processes with careful metadata allow models to learn constitutive behaviour that respects mechanisms rather than forcing fits to superficial curves. Machine-learning surrogates subsequently inherit far less wishful thinking when their training data contain the fingerprints of time, temperature, stress, humidity, and chemistry, recorded with consistent protocols and traceable provenance.

Advanced characterisation bridges simulation and reality by exposing structure and dynamics at the scales where mechanisms live, which matters because many disagreements vanish once both sides can see the same phenomena clearly. Atom probe tomography, aberration-corrected microscopy, operando spectroscopy, scattering under environmental control, and modern tomography make it possible to observe defect populations, grain boundary chemistry, domain structures, surface reconstructions, and transient intermediates as conditions change. When these measurements are planned with model sensitivities in mind, they provide the exact observables required to validate kinetic pathways, parameterise mesoscale models, and test whether a hypothesised mechanism truly governs the behaviour that engineers care about.

Researchers sometimes learn more from deliberately imperfect samples than from carefully polished examples, because robustness reveals itself on the margins where defects, contamination, or process drift begin to matter. Introducing controlled deviations from ideal preparation can quantify how sensitive a mechanism is to manufacturing variability, which helps teams decide whether a candidate warrants investment in process control or whether another candidate offers a wider safe operating window. That information directly affects program risk, because materials that perform acceptably across realistic variation advance more confidently than materials that deliver spectacular numbers only in a narrow laboratory corridor.

Industrial development cares about decision quality more than about elegance, therefore experimentalists increasingly combine measurement with lightweight modelling that translates atomic-scale entitlement into observables that correlate with device-level outcomes. Transfer functions, calibrated on modest datasets, can map simulations to measurable signatures that instrumented platforms can collect quickly, thereby turning expensive calculations into practical screens for ranking candidates and for triaging process options. When these functions carry uncertainty rather than single-point assertions, they support decisions that are accountable to both physics and operations, strengthening trust between computational specialists and teams responsible for production.

Digital twins and virtual reactors become useful once experiments supply the parameters that actually govern behaviour under process constraints, not because a twin replaces measurement, but because it allows researchers to ask counterfactual questions across conditions that would otherwise remain prohibitively expensive. Combining reactive molecular dynamics, mesoscale evolution, and kinetic models with validated parameters enables in silico sweeps that illuminate stability envelopes and failure modes before the first scale-up trial begins. Those insights cut false starts and concentrate experimental effort on the regions where validation has the highest information value for the next design or qualification gate.

The value of national laboratories and coordinated facilities is decisive at this stage because they provide access to high-brightness beams, specialised environments, advanced detectors, and high-performance computing that individual groups cannot sustain alone. These environments also enable reproducibility through standardised workflows, shared ontologies, and data infrastructures where raw measurements, processing metadata, analysis scripts, and uncertainties remain bound together. When users from industry and academia work within these shared platforms, results travel more easily across organisations, and models trained in one context remain interpretable in another because the data carry the context required for honest reuse.

All of this depends on disciplined data practice, because experiments become truly valuable to the broader workflow only when their context survives handoff. Sample history, environment, instrument configuration, calibration references, feature extraction scripts, and uncertainty models must accompany the numbers if we expect any algorithm to generalise responsibly. Without that discipline, machine learning amplifies noise, multiscale workflows fracture at their interfaces, and validation exercises devolve into arguments about provenance rather than arguments about physics. With that discipline, each measured point becomes a durable asset that future programs can mine for insight without reliving the cost of the original experiment.

The practical stance is simple and demanding at once. Treat experiments as instruments for learning rather than as rituals for confirmation, design them to challenge the stories our models prefer to tell, and capture enough context that the resulting data can support models we have not yet written. When computation, machine learning, and measurement work in that spirit, discovery accelerates not because any single component has become miraculous, but because the loop between them has become honest, traceable, and fast enough to keep pace with the complexity of materials that must function reliably in unforgiving environments.

\section{Unlocking the Potential of Quantum Computing}

Quantum computing has drawn enormous attention across materials science because many of the most stubborn problems in the field arise from quantum mechanics itself, and classical approximations begin to fray precisely where the physics grows most interesting. Yet the current hardware landscape forces a more grounded view.~\cite{Ref36_Cao2019, Ref37_McArdle2020} Today's devices remain noisy, limited in scale, and constrained by circuit depth, so near-term value depends on identifying problems where partial quantum information already improves understanding. Many of the hardest materials design questions, however, are likely to require fault-tolerant quantum computing before they can be addressed with confidence.~\cite{Ref38_Preskill2018} For chemically relevant systems, even representing the active space of a modest transition metal complex can already exhaust available qubit counts or circuit depth. It's important to acknowledge that even as the field advances toward fault-tolerant quantum computing, with early fault-tolerant systems at the scale needed to solve real materials problems becoming available in the next 3-5 years, quantum resources will remain scarce and valuable for the foreseeable future. The temptation to oversell quantum possibilities is strong, but the most credible work in the field begins by mapping where the physics genuinely demands a quantum treatment and where classical approaches will continue to lead.

Strong correlation, multi-reference behaviour, and complex spin structures form the core of this territory. Many transition-metal oxides, rare-earth magnets, catalytic active sites, and strongly interacting electron systems fall squarely into regimes where classical methods, even when combined with sophisticated embedding and hybridisation strategies, produce results that require careful interpretation. These systems frequently shape the performance of energy materials, superconducting families, magnetic devices, battery cathodes, and heterogeneous catalysts, and they represent precisely the kinds of problems where quantum solvers might provide an advantage. The goal is not to replace established workflows but to introduce a tool that contributes where classical approximations become unreliable.

Much of the early focus in quantum chemistry has centred on ground-state energies, but energy alone rarely dictates whether a material will succeed in industrial applications. Real value often depends on understanding excited states, reaction pathways, charge migration, optical response, and dynamical behaviour; all of which influence durability, efficiency, selectivity, and safety. Quantum algorithms that treat time evolution natively offer a promising route into these phenomena, because dynamical quantities often defeat classical solvers at scale, yet carry the mechanistic information that guides process design and device performance. When researchers ask which observables matter most for real-world decisions, the list often includes exactly those properties that classical approximations struggle to predict reliably.~\cite{Ref39_HeadMarsden2021}

The difficulty is that quantum devices output only limited information per circuit execution, which forces researchers to think carefully about which observables can be extracted efficiently and which ones require measurement strategies that remain impractical at current noise levels.~\cite{Ref40_Bauer2020} Successful programs start by identifying the variables that carry meaningful physical information without requiring exhaustive reconstruction of the many-body wavefunction. When those variables align with the properties that drive industrial decisions, quantum computing can enter a workflow well before hardware reaches the scale envisioned in long-term roadmaps.

Hybrid strategies are therefore essential, because it is unclear if a quantum device will handle full materials simulations end to end. Classical models still frame the problem, handle large environments, and explore broad composition spaces, while quantum solvers focus on active regions where high accuracy changes conclusions. Embedding schemes that pass information between quantum and classical partitions,~\cite{Ref41_Izsak2022, Ref42_Evenseth2026} must be designed with care, because the limited information content available from quantum measurements places strict constraints on what can be communicated without loss. Nevertheless, these hybrid designs represent the most realistic path to value in the near and medium term, and they naturally evolve as hardware improves.~\cite{Ref37_McArdle2020}

Industrial end-users look at these questions through a pragmatic lens, given the constraints and timeframes they often operate with. A quantum algorithm must do more than produce a theoretically satisfying improvement; it must change the ranking of candidates, shorten development cycles, or reduce the need for costly experiments. Marginal gains in accuracy have little practical effect unless they alter which materials or processes move forward. This requirement forces a focus on efficiency throughout the stack, because any potential advantage evaporates if the overhead of error correction, circuit construction, and measurement eclipses the speed of traditional methods. Industrial researchers increasingly ask targeted questions about where quantum computing can reduce uncertainty in a way that improves yield, stability, or safety, rather than where it can provide a mathematically elegant demonstration disconnected from operational reality.

Simulation of correlated active sites in catalytic cycles, prediction of redox entitlements for strongly interacting electron systems, evaluation of spin-state energetics in rare-earth magnets, and exploration of reaction pathways in electrochemical environments are among the areas where quantum computing could materially shift understanding. These problems routinely require approximations that carry poorly quantified error in classical workflows, and any method that reduces that uncertainty can immediately influence design decisions. The focus therefore shifts from ``quantum advantage'' in the abstract to ``quantum relevance'' in the specific: relevance to design choices, relevance to process windows, and relevance to engineering assumptions that determine whether a material will ever reach production.~\cite{Ref43_Daley2022}

National laboratories and shared research infrastructures play an important role here, because they provide spaces where quantum devices, classical high-performance computing, simulation tools, and advanced characterisation platforms coexist. These environments offer researchers the opportunity to evaluate quantum computing methods under realistic conditions, benchmark them on problems that matter to science and industry, and integrate them into early hybrid workflows without requiring every institution to build a complete quantum ecosystem from scratch. As software stacks evolve and as devices grow more capable, these facilities will shape community norms for testing, validating, and comparing quantum algorithms in ways that reflect genuine materials challenges rather than simplistic demonstrations.

Quantum computing will almost certainly become a hybrid partner rather than a monolithic replacement for classical methods.~\cite{Alexeev2024} Its impact will grow slowly and unevenly, strongest in precisely those pockets of materials science where classical approximations break most dramatically. The challenge for the field lies not only in the development of better quantum algorithms and hardware but in the careful identification of problems where quantum accuracy can reshape decisions. When viewed through that pragmatic lens, quantum computing becomes neither a miracle solution nor a distant aspiration, but a developing capability that can be woven into contemporary workflows as soon as it begins to answer questions that no other method can reach with comparable clarity.

\section{Integrating Quantum and Classical Approaches}

Quantum and classical computing will operate together for a long time because each excels at a different part of the problem and because neither can claim the whole landscape with confidence right now. Classical solvers provide scale, maturity, and broad coverage across structures, temperatures, compositions, and device geometries that industrial programmes must consider when money and time are not theoretical constructs. Quantum solvers promise depth in the small regions where classical approximations start to break down, particularly when correlation, excited states, or nonperturbative dynamics control behaviour in a way that resists tidy treatment.

Integration begins with partitioning, which sounds straightforward and turns out to be demanding once the details arrive. A realistic system must be decomposed so that a quantum region carries the pieces where high accuracy changes the answer, while a classical environment supplies boundary conditions, long-range fields, and the structural context that determine whether the local physics matters in practice. Good partitions respect chemistry and mechanism in the construction of the submodels rather than convenience, so they align with active sites, correlated subspaces, defects, interfaces, or motifs known to steer kinetics and stability under operating conditions. In practice, this often reduces the quantum region to a small, strongly correlated subspace embedded within a much larger environment treated at a lower level of theory.~\cite{Ref44_Thacker2026, Ref45_Montgomery2025}

Information must cross the boundary without losing meaning, which becomes the crux of the method rather than a bookkeeping step that can be postponed. Quantum devices return limited statistics per execution, therefore the interface should carry observables that pack real physical content without requiring state reconstruction that current noise levels cannot support. Energies, reduced density information, response functions, and carefully chosen correlation measures are more helpful than raw bit strings because they speak the same language as the downstream classical models that need to ingest them.

Self-consistency should not be taken for granted, since both sides will feel the influence of the other whenever charge redistribution, spin coupling, strain, or environmental fields play a material role, in particular close to phase transitions. Iterative coupling can stabilise an embedding only when the quantities exchanged are well posed and when the numerical error does not amplify at the interface. Practical workflows therefore stabilise the iterative loop by controlling how updated quantities are introduced between steps, for example by blending new and previous estimates of charge density, potentials, or other exchanged observables such as forces or local potentials so the interface does not oscillate or diverge. Where needed, simplified or surrogate models can be used temporarily to guide convergence before higher-fidelity updates are reapplied.

Many classical workflows were built around classical limitations, which means that inserting quantum outputs as if they were just higher precision numbers will not always work. Some observables that quantum solvers deliver naturally may not map cleanly onto the quantities that classical pipelines expect to process, and some classical simplifications may conceal the very gains that quantum accuracy could provide. Teams should be willing to redesign parts of the pipeline so that quantum-native results can stand as first-class inputs rather than being forced through legacy interfaces that dilute their value.

Decision relevance should guide every integration choice because industrial programmes will not reward elegance that fails to move a gate. Automation then becomes a practical requirement, because manual partitioning will not scale beyond a handful of demonstrations. Sensible heuristics can identify candidate quantum regions by analysing localisation, correlation diagnostics, or the sensitivity of observables to local perturbations, after which human judgment refines the choice. Over time these heuristics should evolve into repeatable workflows that allow non-specialists to apply hybrid methods in a controlled and interpretable way. If a quantum calculation changes a ranking, shrinks an uncertainty interval that blocks a test, or resolves a mechanism that sets a process window, then the integration has justified its place even when the rest of the workflow remains classical. If a calculation produces a finer number that leaves choices unchanged, then the integration has consumed the budget without buying progress.

The orchestration layer benefits from machine learning because a hybrid workflow includes many choices that can be learned from prior runs rather than reinvented for each new system. Surrogates can predict whether a candidate partition will behave well, whether the interface variables will carry enough information, and whether the expected uncertainty reduction justifies the quantum cost at current device performance. Active scheduling can decide which quantum jobs to run next given queue times, error rates, and the marginal value of additional statistics for the specific decision at hand.

Standards and shared infrastructure reduce friction because hybrid methods depend on reliable handoffs between very different tools and facilities. Data formats should preserve provenance, uncertainty measure, and units with enough care that results can be reproduced in another environment without private notes. Reference problems that reflect real materials challenges help the community evaluate methods honestly, since toy models hide the issues that dominate at scale, including memory constraints, measurement budgets, and the tangle of approximations that accumulate across coupled steps.

National laboratories and dedicated institutional ecosystems can shorten the path to deployment by integrating quantum processors directly alongside classical supercomputers and material diagnostics. Within these integrated facilities, users can run end-to-end trials, empirically validate integration architecture, and measure real-world performance gains rather than chasing hypothetical performance metrics. Training programmes that teach engineers how to reason about partitions, observables, and uncertainty will matter as much as device counts, because a capable team extracts more value from modest hardware than an unprepared team extracts from heroic resources.

The hybrid future does not ask researchers to choose sides, because scale and depth are both required for credible materials design. Classical methods will frame the search, provide reach, and deliver the throughput that discovery demands, while quantum methods will target the subspaces where accuracy rewrites conclusions rather than polishing them. The glue that holds the system together will be disciplined interfaces, explicit uncertainty, and workflows that are honest about what each component can and cannot do. When those pieces are in place, integration stops looking like a technology demonstration and starts behaving like a method that teams can trust with decisions that carry real consequences.

\section{Shaping Future Collaboration and Long-Term Research Directions}

\begin{figure*}[t]
    \centering
    \includegraphics[width=0.95\textwidth]{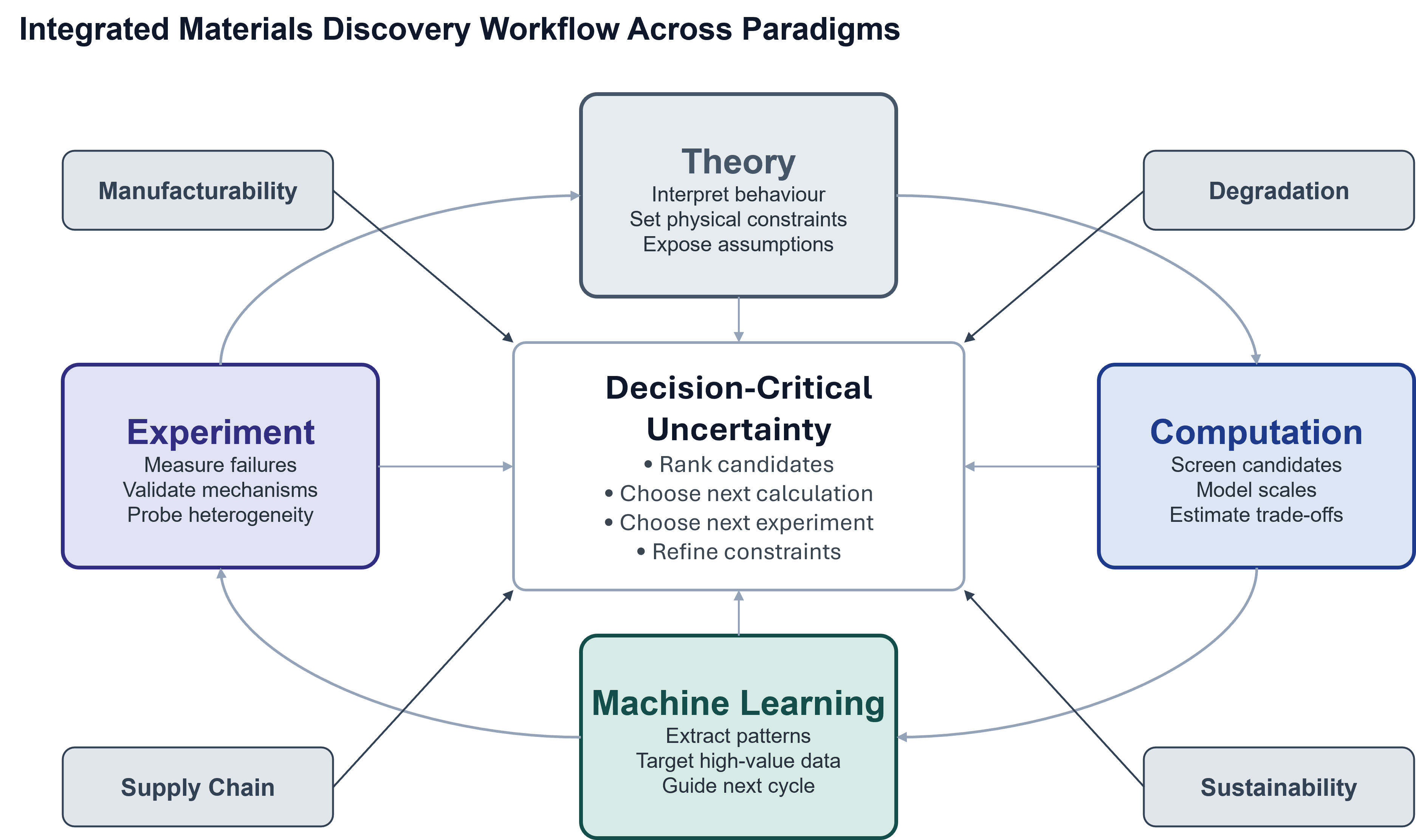}
    \caption{Integration of scientific paradigms within a unified workflow to overcome design and decision-critical constraints in materials discovery.}
    \label{fig:integrated_workflow}
\end{figure*}

Progress in materials discovery increasingly depends on how well different communities work together, not on how elegantly any single group advances its preferred method. As shown in Figure~\ref{fig:integrated_workflow}, integrating multiple paradigms within a unified workflow enables the resolution of design and decision-critical constraints. Universities bring depth and the patience required for foundational work, industry brings the hard edges of manufacturability and economic risk, and national laboratories bring continuity, infrastructure, and the ability to coordinate programs that endure beyond the lifespan of any individual grant. The next decade will be shaped by how deliberately these communities combine their strengths, because the problems that matter most already stretch across time scales, length scales, and institutional boundaries that no single organisation can span alone.~\cite{Ref46_JOM2021}

A durable collaboration begins with shared language rather than shared enthusiasm, since ambiguity about basic terms slows progress more effectively than any hardware limitation. Teams must agree on how uncertainty is represented, how provenance is recorded, and how adequacy is defined for properties that drive engineering choices, otherwise seemingly compatible results cannot be compared honestly. When the vocabulary for error bars, confidence intervals, and validation protocols is aligned across institutions, arguments shift from debating formats to debating physics, which is precisely where disagreements become productive and where the work begins to converge.

Training translators matters as much as training specialists, because the interfaces between disciplines are where ideas either migrate or stall. Graduate programs can help by producing researchers who can read a phase diagram with the same comfort they bring to reading a loss curve, and who can design an experiment informed by a multiscale simulation without treating that simulation as an oracle. Industry can help by supporting secondments that place early-career scientists inside manufacturing environments where constraints become tangible, while laboratories can help by running cohort programs that mix computational scientists, experimentalists, and data engineers on real facility problems rather than on sterile classroom exercises.

Governance structures deserve as much care as algorithms, since collaboration fails quickly when there is confusion about ownership, access, and credit. Pre-competitive spaces that host data models, ontologies, reference workflows, and validation tasks allow competitors to advance the field together without exposing sensitive material, while application-specific models, tuned datasets, and process parameters can remain protected behind well understood contractual boundaries. Attribution must be visible and persistent, because incentives shape behaviour, and because contributors return to ecosystems that recognise their work in durable ways rather than in footnotes that fade once a project moves on.

Shared infrastructure should aim for usefulness rather than spectacle. Data platforms must store raw measurements alongside processing history, analysis scripts, units, and uncertainty, so that a number can be reused responsibly rather than quoted without context. Reference workflows should run end to end with clear versioning, containerised environments, and published test cases that reflect realistic materials problems rather than toy models designed to flatter a single tool. Facilities can publish starter kits that show visitors how to combine characterisation, simulation, and machine learning on a small but complete problem, because a working example carries more persuasive power than a long list of capabilities that remain unconnected in practice.~\cite{Ref47_Jain2013, Ref48_Saal2013, Ref49_Kirklin2015, Ref50_Curtarolo2012}

Industry will continue to ask for demonstrations that move gates rather than for demonstrations that produce attractive figures, which means collaborations should plan explicit decision tests at the outset. If a joint effort claims to reduce uncertainty in a kinetic parameter that governs lifetime, the program should specify which design choice that reduction will unlock, which experiment will validate the claim, and which cost or timeline will change as a result. When success is defined as a decision rather than as a publication, everyone understands why the work matters and where to focus when unexpected complications appear.

National laboratories can play a convening role that others cannot, precisely because they are built to outlast individual projects and because they operate facilities that anchor large communities. They can host hybrid testbeds where quantum devices, high-performance classical computing, and advanced characterisation operate in coordinated workflows, and they can maintain benchmark problems that evolve as the field learns rather than as the field markets itself. They can also sustain the unglamorous work of software modernisation, metadata standards, and data stewardship, which quietly determine whether progress accumulates or whether each group repeats the same hard lessons in isolation.

Academic groups can preserve the freedom to chase hard questions while still contributing to shared foundations that serve the broader community. Method development gains traction when released with clear regimes of applicability, reproducible notebooks, and pathways for integration into multiscale workflows rather than with isolated code that remains impossible to adopt. Students who learn to publish along with environments, test cases, and data citations carry habits that make their work transfer effectively, which benefits both the academy and the employers who later rely on those graduates to anchor industrial programs.

Not every collaboration needs to scale to a consortium, but not every problem can be solved by a small team either. Programmes that require tight integration across experiment, computation, and theory, particularly when they involve shared infrastructure or multiscale workflows, often benefit from larger, coordinated groups with clearly defined interfaces. At the same time, focused problems with well-defined objectives, such as method development or targeted validation, often move faster in smaller teams where feedback loops are shorter and decisions can be made without organisational overhead. The challenge is not choosing one model over the other, but matching the scale of collaboration to the structure of the problem.

Sustainability and resilience need to appear as first-class objectives inside collaborations, not as late appendices that mention compliance in passing. Materials that look promising on performance alone often falter once lifecycle, criticality, recyclability, and regulatory limits are applied with the same seriousness that performance receives, and collaborations that ignore these realities waste time on candidates that will never reach a product line. When these considerations are encoded inside search, modelling, and experimental design from the start, programs move faster, because they stop promoting options that careful analysis would have filtered at the first gate.

Overall, the field will benefit from a cultural shift that treats infrastructure building as scientific work rather than as support work. People who design databases that actually capture process history, who write interfaces that let quantum and classical solvers exchange meaningful observables, who curate datasets that stand up to unintended reuse, and who maintain reference workflows across versions are doing research that changes what becomes possible next year. Recognising these contributions with authorship, funding paths, and career advancement will accelerate progress more reliably than any announcement about a single breakthrough, because it strengthens the soil in which many breakthroughs can take root.

If collaborations are built on shared language, clear governance, honest benchmarks, and training that values translators as much as specialists, the next decade can move beyond demonstrations that impress in isolation. The community can then deliver workflows that survive contact with industry, with regulation, and with the messy realities of scale, which is where scientific ideas either become technologies or remain attractive stories that never quite reach the world that needs them.

\section{What the Field Needs Next}

The tools that shape today’s materials research are already powerful enough to support significant advances, yet the field often struggles to turn these isolated successes into capabilities that can be used reliably across different organisations and scientific cultures. The next phase of progress will depend on whether researchers can create an environment where knowledge, data, and workflows move cleanly between teams, allowing strengths from one domain to reinforce strengths from another. This requires a deliberate shift in how the community handles uncertainty, curates information, trains future scientists, and recognises the work that makes collaboration possible. Without these foundations, even the most sophisticated ideas lose momentum once they leave the controlled setting in which they were created.

Uncertainty will remain a defining feature of materials discovery, and the field needs conventions that treat it as something to be examined instead of something to be hidden. Every computational technique, whether it involves electronic structure, atomistic simulation, mesoscale modelling, or machine learning, produces results that depend on assumptions about structure, sampling, and numerical choices. Those assumptions shape the reliability of the predictions far more than the final numbers suggest, yet the associated uncertainty often disappears by the time conclusions are presented. A healthier scientific ecosystem would make uncertainty visible at each step, tying it directly to the decisions it influences. When teams see precisely which parts of a workflow carry the most fragility, they can focus their efforts on the experiments or higher‑fidelity simulations that reduce the risk of costly misdirection.

Progress also depends on consistent data practices that reflect the messy reality of materials research rather than the convenience of individual groups. Many of the properties that guide technological decisions emerge slowly through degradation, fatigue, creep, corrosion, or chemical drift, and these behaviours depend critically on sample preparation, processing routes, environmental conditions, and instrument settings. Without careful capture of these details, datasets become brittle, misleading, or impossible to reuse. A shared schema that stores values with complete provenance, explicit uncertainty, and a faithful record of experimental or computational history would transform how teams evaluate prior results.~\cite{Ref51_Wilkinson2016, Ref52_Draxl2018} Such a framework would allow researchers to combine insights across institutions, compare workflows honestly, and avoid the duplication of studies that could have been resolved with responsibly curated data.

Training must evolve toward a model that prepares scientists to navigate the increasingly interconnected nature of the field. Materials research no longer unfolds within the boundaries of a single discipline, and the most challenging problems demand fluency across computation, experiment, data science, and engineering constraints. Students should learn to interpret simulation outputs through the lens of their approximations, to design experiments that confront the weaknesses of a model rather than the comforts of a hypothesis, and to recognise when a machine‑learning trend reflects physics rather than artefacts of sampling. This broader literacy will not replace deep expertise, but it will equip researchers with the judgment required to move ideas across the conceptual boundaries that once separated the major paradigms of the field.

Infrastructure deserves equal attention, because the systems that store data, manage workflows, preserve provenance, and coordinate computational resources determine whether knowledge endures or dissipates. Databases that maintain raw data alongside detailed metadata, workflow engines that guarantee reproducibility across environments, and interfaces that enable information to pass between quantum solvers, classical models, and machine‑learning tools form the backbone of modern materials research. People who build and maintain these systems make scientific progress possible, and the field must recognise that their work carries intellectual value rather than merely operational value. When infrastructure is treated as a core scientific product, it becomes easier for organisations to share tools, adopt best practices, and build upon each other's progress instead of repeatedly reinventing the same foundations.

Sustainability must shape research directions at the earliest stages instead of appearing as a constraint that becomes visible only after performance metrics have already captured attention. Materials that excel in simulations or small‑scale experiments can become untenable once their resource demands, environmental impacts, or regulatory limitations emerge with full force. Aligning discovery strategies with lifecycle considerations, supply chain resilience, recyclability, and environmental boundaries will reduce the number of candidates that falter late in development. It will also ensure that promising pathways receive attention early enough to influence how tools, models, and characterisation methods evolve.

Collaboration across sectors will be essential for the field to mature. Universities contribute fundamental understanding and the curiosity that drives conceptual advances, industry contributes constraints that reflect real stakes and real timelines, and national laboratories contribute the infrastructure that anchors long‑term scientific programmes. These groups can deliver far more together than they can separately, but only when the rules governing ownership, access, and attribution are clear. When data can move across institutions without confusion, when credit aligns with actual contributions, and when incentives encourage cooperation instead of competition in foundational areas, the field becomes capable of solving problems that no single organisation could attempt alone.

Realistic benchmarking will help guide these efforts. Many methods achieve impressive results on narrow test sets that conceal the complexities of genuine materials behaviour. Models should be evaluated on challenges that reflect the disorder, correlation, environmental variability, and long‑term evolution that determine success in industrial settings. Benchmarks that reward interpretability, robust uncertainty handling, physical validity, and reproducibility will encourage the development of methods that help researchers make real decisions rather than methods that win artificially simplified competitions.

Finally, the field must invest in the long‑term preservation of knowledge.~\cite{Ref53_Oviedo2022} Materials discovery often cycles through ideas that reappear every decade, sometimes without awareness of older work that would have prevented unnecessary detours. Curated datasets, versioned workflows, open protocols, and shared repositories become the collective memory that allows new projects to begin from a higher starting point. When knowledge is preserved with care, progress becomes cumulative rather than cyclical, and scientific energy can be directed toward new questions rather than toward rediscovering what was once understood.

If these structural needs are met with seriousness and shared commitment, the field will move from incremental steps to genuine acceleration. Scientific tools already exist, but their impact depends on the systems that surround them, the people who connect them, and the culture that encourages openness, rigor, and collaboration. When those conditions take hold, the path from insight to application becomes clearer, faster, and more resilient, allowing the community to tackle problems whose complexity once placed them beyond reach.

\section*{Conclusions}

Materials discovery has entered a period where progress depends less on isolated advances and more on the coherence of the system that surrounds them. The field now possesses computational methods capable of exploring enormous design spaces, experimental platforms capable of revealing behaviour across scales and environments, and data-driven techniques capable of extracting structure from the complexity that defines real materials. These tools allow researchers to imagine possibilities that would have been unreachable even a decade ago, yet their full potential remains constrained by fragmentation, uneven data practices, and a persistent disconnect between idealised workflows and the realities of scale, cost, and uncertainty. The next phase of progress will come from the bridges that connect these capabilities rather than from the individual capabilities themselves.

A common thread runs through the scientific challenges described in this manuscript. Whether the focus involves strong correlation, microstructural evolution, kinetic pathways, degradation mechanisms, manufacturability constraints, supply chain stability, or sustainability demands, the most important insights emerge when computation, experiment, and machine learning operate as parts of a single reasoning process rather than as independent activities. When the information that flows through a workflow carries its provenance, uncertainty, and mechanistic context, researchers can identify which directions deserve investment, which approximations remain defensible, and which apparent signals represent artefacts of sampling rather than the physics of the system. Decisions improve when workflows become honest about their own limits, and innovation accelerates when those limits encourage new measurements, new models, or new conceptual frameworks.

The scientific landscape is too broad and too intricate for any single group to master, and the field will advance fastest when universities, industry, and national laboratories operate as complementary contributors to a shared ecosystem. Fundamental insights from academia gain power when translated through industrial constraints that reveal what performance truly requires, while laboratory facilities provide the continuity, infrastructure, and methodological discipline that allow ideas to mature. These partnerships become durable only when the underlying governance, attribution, and data practices support trust rather than friction, and when each institution recognises the value that others bring to problems that no isolated team can solve alone.

Sustainability, resilience, and societal responsibility should steer discovery with the same influence as performance metrics, because materials that cannot be manufactured at scale, recovered responsibly, or aligned with regulatory boundaries rarely deliver value beyond the laboratory. Embedding these considerations into the earliest stages of modelling, screening, and experimentation will help the field avoid pathways that sparkle briefly before revealing fundamental incompatibilities with real-world constraints. This shift encourages researchers to search for solutions that perform well while also surviving the broader pressures of a world that increasingly demands responsible innovation.

Knowledge preservation will also shape the trajectory of the field. As long as data, models, and workflows remain locked inside individual projects or institutions, progress will continue in cycles that rediscover old dead ends and repeat avoidable mistakes. When the community invests in infrastructure that stores knowledge with transparency and care, future researchers begin their work from a higher platform. Their attention can move toward unanswered questions rather than toward reconstructing information that once existed but was never captured with the fidelity required for reuse.

If the field embraces these structural needs with intention and clarity, the coming years can yield a level of acceleration that surpasses what any single technological advance could deliver on its own. The scientific tools now available are already formidable, yet their greatest impact will emerge only when they are woven into workflows that respect the complexity of materials and the constraints of the world in which those materials must operate. A future defined by integrated reasoning, reproducible practice, shared infrastructure, responsible data stewardship, and collaborative problem solving will allow the community to approach materials design with a confidence and pace that were previously out of reach.

When the conversation moves fluidly between computation, experiment, machine learning and artificial intelligence, quantum methods, engineering realities, and societal considerations, materials science becomes capable of addressing challenges whose complexity once placed them beyond credible reach. The goal is not simply to discover new materials but to discover materials that endure the full journey from conceptual possibility to real-world function, and to ensure that the discoveries made today become the foundations on which future generations can build with clarity, stability, and imagination.

\section*{Author contributions}

The ideas presented in this manuscript grew from sustained collaboration between researchers working in academia, industry, and national laboratories. Each author contributed perspectives grounded in their own experience, and the final narrative reflects the integration of these viewpoints into a coherent argument about the future of materials discovery. Although the writing has been synthesised into a unified voice, the intellectual contributions remain rooted in the expertise that each participant brought to the project.

\vspace{0.5em}

\noindent \textbf{Phalgun Lolur:} Project administration, Conceptualization, Writing -- original draft, Writing -- review \& editing. Coordinated the development of the manuscript, shaped the overarching structure and contents, integrated contributions into a unified narrative, and led the synthesis across computation, experiment, data-driven methods, and quantum technologies. Worked with contributors to refine arguments, ensure conceptual continuity, and align the manuscript with scientific priorities.

\vspace{0.5em}

\noindent \textbf{Richard Padbury:} Conceptualization, Writing -- review \& editing. Co-developed the original framing of the manuscript and contributed to discussions that shaped the central questions addressed across sections. Influenced the conceptual structure of the article and helped establish the themes connecting academic, industrial, and national laboratory perspectives.

\vspace{0.5em}

\noindent \textbf{George H. Booth:} Conceptualization, Writing -- review \& editing. Provided foundational insights into method adequacy, uncertainty propagation, electronic structure limitations, and quantum relevance. Shaped the sections on the four scientific paradigms, computational methods, and quantum algorithms, particularly regarding dynamical observables and multi-method consistency.

\vspace{0.5em}

\noindent \textbf{Katherine Inzani:} Conceptualization, Writing -- review \& editing. Contributed perspectives on research cultures, training models, cross-paradigm literacy, and the evolving landscape of collaborative science. Influenced the sections on paradigm integration, education, and the structural needs that support sustained scientific progress.

\vspace{0.5em}

\noindent \textbf{Nicole Holzmann:} Conceptualization, Writing -- review \& editing. Contributed viewpoints on quantum computing that emphasise stack efficiency, property-driven value, and workflow design that respects the realities of industrial decision making. Informed the quantum sections and the supplementary material focused on responsible integration of quantum methods.

\vspace{0.5em}

\noindent \textbf{Thomas W. Keal:} Conceptualization, Writing -- review \& editing. Contributed to discussions on national laboratory infrastructure, exascale readiness, multiscale approaches and their integration with data-driven methods and quantum computing, and the reproducibility challenges that shape community practice.

\vspace{0.5em}

\noindent \textbf{Joseph Montoya:} Conceptualization, Writing -- review \& editing. Contributed perspectives on industrial modelling practice, trend-guided computation, disorder-dominated behaviour, and virtual reactor methods. Shaped the discussion of early-stage discovery, computational realism, and small-data decision making.

\vspace{0.5em}

\noindent \textbf{Daniel F. Urban:} Conceptualization, Writing -- review \& editing. Contributed expertise on microstructure-driven properties, rare events, multiscale modelling, and the needs of long-term materials reliability. Helped shape the sections on physical experimentation, multiscale simulation, and national laboratory workflows.

\vspace{0.5em}

\noindent \textbf{Thomas Eckl:} Conceptualization, Writing -- review \& editing. Provided industrial insights on manufacturability, supply chain stability, safety constraints, and concurrent engineering. Strengthened the arguments for embedding real-world constraints directly into optimisation and reinforced the importance of data architecture that captures uncertainty and context.

\vspace{0.5em}

\noindent \textbf{Emanuele Marsili:} Conceptualization, Writing -- review \& editing. Contributed viewpoints on multi-objective design, corrosion modelling, environmental dependence, and the integration of atomistic insights into engineering decisions. Influenced discussions on the four paradigms, computational methods, and industry-aligned discovery workflows.

\vspace{0.5em}

\noindent \textbf{Wibe de Jong:} Conceptualization, Supervision, Writing -- review \& editing. Provided guidance on computational workflows, national laboratory perspectives, and community-scale coordination across software, data, and training. Influenced sections involving infrastructure, reproducibility, and the role of shared platforms.

\vspace{0.5em}

\noindent \textbf{Jonathan R. Owens:} Conceptualization, Writing -- review \& editing. Contributed perspectives on the integration of computational methods with experimental validation in industrial materials discovery, with a particular emphasis on translating atomic-scale insights into decision-relevant outcomes. Provided input on workflow design, multiscale simulation strategies, and the role of hybrid approaches in addressing complex materials challenges in energy and advanced engineering systems.

\vspace{0.5em}

\noindent \textbf{Julian van Velzen:} Conceptualization, Writing -- review \& editing. Contributed perspectives on lifecycle constraints, sustainability, resource criticality, and atomic-scale fidelity for systems near their reliability limits. Informed the sections on early discovery, industrial needs, and sustainability-aligned design.

\section*{Conflicts of interest}

There are no conflicts of interest to declare. Dr. Richard Padbury contributed to this work as an independent researcher; the views expressed herein are entirely his own and do not represent those of his current or former employers.

\section*{Data availability}

No primary research results, software or code have been included and no new data were generated or analysed as part of this review.



\section*{Author Biographies}

\noindent \textbf{Prof. George Booth} is a theoretical physicist at King's College London, U.K. His interests lie primarily in computational electronic structure theory, with a particular focus on the development of novel computational techniques that push the state of the art in both classical and quantum algorithms in scope, scale, and accuracy. With a background encompassing both chemistry and physics perspectives, he has a particular interest in the development of scalable approaches for extended systems and stochastic methods, working at the intersection of traditional quantum chemistry and numerical physics approaches across condensed matter, materials science, and chemical application domains.

\vspace{1em}

\noindent \textbf{Dr. Wibe de Jong} serves as the Department Head for Computational Sciences, and leads the Applied Computing for Scientific Discovery Group, which advances scientific computing by developing and enhancing applications in key disciplines, as well as developing tools and libraries for addressing general problems in computational science. Dr. de Jong is the Director of the Quantum Systems Accelerator, which is part of the National Quantum Initiative. In addition, de Jong is the Team Director of the Accelerated Research for Quantum Computing (ARQC) Team MACH-Q, funded by DOE ASCR, focused on developing error-aware software stacks for hybrid quantum computing devices. He is a co-PI on two science with quantum focused projects in DOE Basic Energy Sciences and DOE High-energy Physics.

\vspace{1em}

\noindent \textbf{Dr. Thomas Eckl} is Chief Expert for Computational Materials Design at Robert Bosch GmbH in Stuttgart, where he works within Bosch Corporate Sector Research and Advance Engineering. His work focuses on the simulation and design of advanced materials for batteries, fuel cells, e-motors, and sensors using computational methods---an approach that complements experimental development and accelerates innovation in industrial applications. He is an active contributor to the field of computational materials science, particularly in the application of quantum and classical simulation techniques to understand complex materials systems like strongly correlated transition metal oxides. In recent years, his work has increasingly explored the use of hybrid quantum--classical algorithms and quantum computing to overcome the limitations of conventional simulation approaches, especially for materials with complex electronic interactions. 

\vspace{1em}

\noindent \textbf{Dr. Nicole Holzmann} is the Senior Applications Lead --- Defense at PsiQuantum, where she identifies and develops viable use cases for fault-tolerant quantum computing in defense and aerospace. Previously at Riverlane, Nicole led a team evaluating quantum algorithms and their practical feasibility across industries. With over a decade of academic research in quantum and computational chemistry, covering catalysis, membrane proteins, and drug design, Nicole bridges deep technical expertise with real-world application development, helping close the gap between quantum computing's promise and its practical impact.

\vspace{1em}

\noindent \textbf{Dr. Katherine Inzani} is an Associate Professor in Chemistry at the University of Nottingham, UK, where she leads research in computational materials design for next-generation electronic and quantum technologies. Her work focuses on developing theory-led approaches to uncover atomic-scale mechanisms governing functional materials, with a particular emphasis on linking these insights to device-relevant performance. She holds a UKRI Quantum Technologies Fellowship, where her work focuses on embedding fundamental materials research within application-driven contexts. A particular interest lies in integrating first-principles approaches with machine learning methods and quantum computing to enable predictive, scalable materials design.

\vspace{1em}

\noindent \textbf{Prof. Thomas W. Keal} leads the Multiscale Chemistry Group at STFC Scientific Computing, based at Daresbury Laboratory in the UK, and is a visiting professor at the University College London Department of Chemistry. His research interests are in the development of hybrid quantum mechanics/molecular mechanics (QM/MM) approaches for the simulation of complex chemical, biomolecular, and materials systems, and he leads the development of the ChemShell multiscale computational chemistry environment, a leading software package for QM/MM simulations. Recent projects have included the development of multiscale workflows for exascale HPC platforms and multiscale modelling frameworks that integrate HPC and quantum computing algorithms.  

\vspace{1em}

\noindent \textbf{Dr. Phalgun Lolur}  serves as the Head Scientist of the Capgemini Quantum Lab, leading technical development at the intersection of quantum chemistry, materials science, data-driven methods, and quantum computing. Endorsed by the Royal Society for his expertise in theoretical, computational, and quantum chemistry, he brings over 16 years of experience in quantum simulations, optimisation, and machine learning. In his current role, he oversees multidisciplinary teams to develop solutions for complex challenges in the life sciences, chemicals, materials, and energy sectors, with a focus on structuring how quantum, AI, and classical methods combine within end-to-end workflows. His work centres on developing multi-scale modelling frameworks that connect atomistic simulations with real-world operating conditions, bridging the gap between theoretical models and deployable applications.

\vspace{1em}

\noindent \textbf{Dr. Joseph Montoya} is a senior staff research scientist in the Energy \& Materials division at Toyota Research Institute (TRI). Dr. Montoya specializes in designing software for the discovery of new materials. His research interests include catalysis, reaction engineering, solid mechanics, and corrosion engineering using density functional theory calculations. Prior to joining TRI, Dr. Montoya earned his Ph.D. in chemical engineering at Stanford University ('15) and was a postdoctoral fellow with the Materials Project at Lawrence Berkeley National Laboratory, where he developed tools for the automation of high-throughput simulations, dissemination of data on the web, and analysis of materials data for the prediction of novel materials.

\vspace{1em}

\noindent \textbf{Dr. Jonathan R. Owens} works at the intersection between computational physics, chemistry, and materials science at the GE Vernova Advanced Research Center in Niskayuna, NY. His research interests include applications of computational methods for material discovery and characterization, including \ce{CO2} and \ce{H2O} sorbents, thermal barrier coatings, thermoelectric materials, and magnetic materials. He is also interested in emerging techniques to enable more accurate computational approaches, such as machine learning and quantum computing, and how those technologies can be used for accelerated industrial materials design. He also holds a Visiting Scientist appointment at the Ralph O'Connor Sustainable Energy Institute at Johns Hopkins University.

\vspace{1em}

\noindent \textbf{Dr. Richard Padbury} is a research executive and strategist specializing in AI, data, and emerging technologies. With a strong emphasis on cross-functional exploration, his work focuses on applied research strategy, emerging technology evaluation, and fostering collaboration across industry, academia, and technology partners. Notably, he has led materials science research programs designed to bring experimental and computational workflows together to solve complex challenges in materials development and formulation. Additionally, he has published research in quantum algorithms, including work in quantum game theory, Hamiltonian simulation, and quantum graph neural networks. He holds a B.Sc. in Physics from the University of Manchester (UK) and a Ph.D. in Polymer Science from NC State University. 

\vspace{1em}

\noindent \textbf{Dr. Daniel F. Urban} heads the materials modelling group at the Fraunhofer Institute for Mechanics of Materials IWM in Freiburg, Germany. Specializing in the physical modelling and numerical simulation of materials on the atomic scale, his research contributes to the understanding of structure-composition-property relationships of various functional materials, including amorphous transparent conducting oxides, high-performance permanent magnets, Li-ion conductors, and high-temperature corrosion-resistant steels. A current focus of his research is on the modelling of highly correlated materials relevant for energy storage and conversion, as well as the development of hybrid classical/quantum computing approaches for calculating their electronic structure.

\vspace{1em}

\noindent \textbf{Julian van Velzen} heads Capgemini's Quantum Lab. With a background in condensed matter physics, his work focuses on translating deep-tech capabilities into credible industrial and societal impact. He oversees internal research programs, client engagements, and capability development spanning areas such as quantum optimisation, chemistry, materials science, and quantum-safe technologies. He is actively involved in building international research and partner ecosystems, and contributes to shaping how emerging quantum technologies can be responsibly integrated into real-world decision-making and large-scale engineering workflows.


\bibliographystyle{unsrt}  
\bibliography{references}

\newpage
\appendix
\section{Supplementary Information}
\label{sec:supplementary_material}

\renewcommand{\thefigure}{S\arabic{figure}}
\renewcommand{\thetable}{S\arabic{table}}
\renewcommand{\theequation}{S\arabic{equation}}
\setcounter{figure}{0}
\setcounter{table}{0}
\setcounter{equation}{0}

\subsection{Introductory note and reading guidance}

This section gathers additional perspectives that informed the manuscript yet did not fit cleanly inside the main arc without interrupting the flow. The contributions retain the authors' original stance and emphasis, while the language has been lightly shaped so the ideas remain compatible with the narrative voice used throughout the article. Each perspective should be read as a lens rather than as a standalone argument, since the value comes from how these lenses help the reader interrogate methods, recognise limits, and choose the next experiment or computation with sharper intent.

Readers who want to apply these insights in practice can use this section as an index of nuance that complements the core sections. If a pipeline struggles with uncertainty propagation across scales, the entries that stress methodological adequacy and cross-validation will help. If a team needs to integrate manufacturability and lifecycle constraints earlier in the search, the entries that foreground enterprise realities will provide tangible direction. If a programme aims to evaluate quantum computing with industrial discipline, the entries that frame stack efficiency, property relevance, and interface variables will save time and prevent avoidable detours.

\subsection{Academic perspectives and method foundations}

\noindent\textbf{George H. Booth:} \textit{capability maps, uncertainty that travels, and quantum targets that actually matter}

\hangindent=0.5cm\hangafter=0\small\noindent
George stresses that reliable discovery depends on knowing where methods work, where they fail, and how errors move through the pipeline when models hand results to one another. The field needs capability maps by property and by materials class, together with uncertainty that accompanies a quantity from electronic structure through interatomic models into mesoscale and device levels without evaporating during post-processing. Methodological diversity prevents pipelines from inheriting the blind spots of a single approach, while honest escalation rules ensure that higher fidelity appears when lower levels begin to drift. On the quantum side, he argues that clearest long-term value lies with dynamical observables and higher-order response rather than with static energetics alone, because time evolution and strongly correlated dynamics defeat classical solvers at scale and carry the kind of mechanistic information that changes engineering decisions.\par

\vspace{1.5em}

\noindent\textbf{Katherine Inzani:} \textit{training translators, integrating paradigms, and balancing exploration with exploitation}

\hangindent=0.5cm\hangafter=0\small\noindent
Katherine highlights that the most productive research cultures train people to move across empirical, theoretical, computational, and data-driven paradigms without treating any of them as an oracle. She points to cohort models, cross-disciplinary consortia, and curriculum designs that prepare students to translate between methods, ask integrative questions, and design experiments that probe model sensitivities rather than confirm expectations. In early discovery she emphasises structured search, where active learning, evolutionary strategies, and Bayesian optimisation help balance exploration of unfamiliar regions with exploitation of promising leads, while dimensionality reduction and physically motivated descriptors make high-dimensional spaces navigable without losing mechanistic meaning. She also reminds the community that some problems need better data and integration rather than new algorithms, while others genuinely require foundational method work.\par

\subsection{Industry perspectives and enterprise constraints}

\noindent\textbf{Julian van Velzen:} \textit{lifecycle constraints, atomic-scale fidelity where reliability lives, and access through modern infrastructure}

\hangindent=0.5cm\hangafter=0\small\noindent
Julian calls for lifecycle and resource constraints to appear inside early screening so candidate sets survive contact with sustainability targets and supply realities. He notes that many high-value systems operate near reliability limits where atomic-scale effects dominate behaviour, which makes quantum-accurate modelling an essential frontier for certain classes of problems. He also observes that cloud platforms and modern accelerators have lowered the barrier to high-fidelity modelling, allowing more teams to use advanced electronic-structure methods and curated datasets responsibly, provided they adopt standards that preserve provenance, uncertainty, and interoperability.\par

\vspace{1.5em}

\noindent\textbf{Emanuele Marsili:} \textit{engineered trade-offs, corrosion realism, and the right way to pair generative models with physics}

\hangindent=0.5cm\hangafter=0\small\noindent
Emanuele brings an engineer's view of multi-objective design where performance must coexist with safety, stability, process windows, and cost. He argues for formal optimisation frameworks that capture those trade-offs directly, since ad hoc checks hide assumptions and delay unwelcome truths. In corrosion and inhibitor design he emphasises environmental dependence, with parameters that move under changes in humidity, temperature, electrolyte composition, and surface condition, which makes atomistic inputs to mesoscale and continuum models particularly valuable. He treats generative models as productive only when coupled to simulations or targeted measurements that provide dynamic labels, because static annotations rarely capture behaviour in mixtures, at interfaces, or under service environments that matter most to industry.\par

\vspace{1.5em}

\noindent\textbf{Joseph Montoya:} \textit{computation as a compass, disorder as a fact of life, and virtual reactors that respect process knobs}

\hangindent=0.5cm\hangafter=0\small\noindent
Joseph frames computation as guidance rather than as a promise of one-to-one mapping, especially in regimes where disorder, amorphous environments, or complex surfaces dominate the measured response. He points to catalytic trends such as relationships between electronic descriptors and adsorption behaviour as examples of how approximate models can still direct formulation and screening. He also advocates for virtual reactors that represent process variables like temperature, pressure, gas composition, flow, mixing, and time profiles, since industrial levers rarely match the boundary conditions assumed in clean theoretical treatments.\par

\vspace{1.5em}

\noindent\textbf{Jonathan R. Owens:} \textit{entitlement with accountability, small-data realism, and industrial thresholds for quantum value}

\hangindent=0.5cm\hangafter=0\small\noindent
Jonathan underscores the value of using simulation to establish entitlement while respecting the limits imposed by defects, disorder, and macro-scale behaviour. He recommends lightweight transfer functions that map atomic-scale predictions to measured observables when sufficient data exist, with uncertainty carried transparently so engineers can act with confidence. In small-data regimes he favours classical approaches such as kernel methods and Gaussian processes because they provide calibrated uncertainty and disciplined behaviour. He views quantum adoption through a pragmatic lens, where value appears only when a calculation becomes cheaper and faster than the corresponding experiment for properties that drive real industrial choices.\par

\subsection{National laboratory perspectives and shared infrastructure}

\noindent\textbf{Daniel F. Urban:} \textit{microstructure, rare events, and community standards that make multiscale models honest}

\hangindent=0.5cm\hangafter=0\small\noindent
Daniel focuses on aspects that determine reliability and lifetime of materials and devices, such as crack initiation, fatigue, corrosion, and microstructural evolution across long timescales. He calls for multiscale models that pass physically meaningful state variables rather than only effective numerical parameters, so interpretability survives the journey from atoms to devices. He stresses that strong ontologies, measurement standards, and FAIR principles must cover experiments as carefully as simulations, because lifetime data lose most of their value when context and uncertainty fail to travel with the numbers. He also treats quantum computing as potentially useful for classes of correlated problems, while reminding the community that value depends on resources, interfaces, and integration rather than on feasibility alone.\par

\vspace{1.5em}

\noindent\textbf{Thomas Keal:} \textit{modernising codes, embedding as a pattern, and facilities as engines for reproducibility}

\hangindent=0.5cm\hangafter=0\small\noindent
Tom highlights the work required to adapt large community codes to heterogeneous hardware and exascale architectures, since software that cannot evolve will limit scientific reach regardless of algorithmic promise. He views embedding approaches as a reusable pattern for hybrid workflows, where high-accuracy regions can expand as hardware and algorithms improve without forcing complete redesigns, and are well placed to adapt to incorporate new technologies including machine learning methods and quantum computing. He emphasises the role of facilities in providing shared experimental and computational environments, training, benchmark problems, and reproducible stacks that allow methods to be compared on tasks that reflect real world materials problems rather than curated demonstrations.\par

\subsection{Cross-functional perspective on workflows and data architecture}

\noindent\textbf{Thomas Eckl:} \textit{engineering constraints as design parameters, concurrent loops, and data backbones that prevent fragile optimisation}

\hangindent=0.5cm\hangafter=0\small\noindent
Thomas argues that manufacturability, supply chain stability, safety constraints, and regulated substances belong inside optimisation rather than outside it, because the most damaging failures occur when late filters overturn decisions that looked attractive earlier. He promotes concurrent loops where empirical testing, theory, automated simulation chains, and data-driven surrogates operate together, guided by uncertainty rather than by habit. He calls for data architecture that enforces standard formats, strong metadata, and explicit uncertainty quantification so models do not become brittle when they face real world validation. Across scales he urges the passage of physically meaningful variables rather than only abstract fitted parameters, since interpretability and trust rise when mechanisms remain visible during upscaling.\par

\subsection{Quantum value, stack efficiency, and workflow redesign}

\noindent\textbf{Nicole Holzmann:} \textit{economic meaning, observables that drive decisions, and pipelines that accept quantum-native results}

\hangindent=0.5cm\hangafter=0\small\noindent
Nicole frames quantum advantage as economically meaningful only when accuracy and cost combine to change an engineering choice, a development timeline, or a risk posture. She points to stack efficiency as a non-negotiable requirement, because any theoretical speedup evaporates when error correction, circuit depth, and measurement overhead dominate runtime and budget. She also notes that many quantum outputs do not map directly onto the observables that industrial teams need, which means workflows should be redesigned so quantum-native quantities can stand as first-class inputs rather than being squeezed through classical interfaces that were never built for them.\par

\vspace{1.5em}

\subsection{How to use these perspectives in practice}

Readers can treat this section as a practical companion to the main text. When a workflow needs disciplined uncertainty that moves across scales, the entries by George and Daniel provide concrete direction. When a programme requires search strategies that keep exploration honest while moving efficiently, the entries by Katherine and Joseph offer guidance that connects algorithms to mechanisms. When enterprise constraints must shape the search space from the start, the entries by Kristin, Julian, and Thomas provide the language and structure needed to encode those realities directly into optimisation. When an organisation wants to evaluate quantum computing without theatre, the entries by Nicole, George, Tom, and Daniel explain where value can appear, how interfaces must behave, and which observables actually matter. When teams struggle to make results travel, the entries across the national laboratory perspectives show how standards, modernised codes, shared environments, and reproducible workflows convert isolated successes into durable capability.

This supplemental material exists to preserve nuance without cluttering the main narrative, so teams can move from ideas to decisions with fewer detours. If a reader finds that a difficult choice depends on a detail that the core sections did not expand, the chances are good that one of these perspectives has already explored that corner of the landscape and has left a navigational clue that will repay careful attention.

\end{document}